\documentclass[prb,twocolumn]{revtex4}
\input epsf

\begin{document}

\title{
 Relativistic Causality and
  Conservation of Energy \\  in Classical Electromagnetic Theory}

\author{A. Kislev}
\affiliation{Iscal Holdings \\6 Shoham Street, Petach Tikva 49517, Israel}

\author{L. Vaidman}

\affiliation{  Centre for Quantum Computation\\
 Department of Physics, University of Oxford,\\
Clarendon Laboratory, Parks Road, Oxford OX1 3PU, England}

\affiliation{
 School of Physics and Astronomy\\ 
Raymond and Beverly Sackler Faculty of Exact Sciences \\
Tel-Aviv University, Tel-Aviv 69978, Israel}

\date{}

\vspace{.4cm}
\begin{abstract}
  The change of the electromagnetic field in a particular place due to
  the event of a change in the motion of a charged particle can occur
  only after the light signal from the event can reach this place.
  Naive calculations of the electromagnetic energy and the work
  performed by the electromagnetic fields might lead to paradoxes of
  apparent non-conservation of energy. A few paradoxes of this type for
  a simple motion of two charges  are presented and
  resolved in a quantitative way providing deeper insight into various
  relativistic effects in the classical electromagnetic theory.
\end{abstract}
\vspace {.5cm}

\maketitle

\section{INTRODUCTION}
\label{int}

Starting from Einstein's work on special relativity \cite{Eins} it
became clear that classical electromagnetic theory is consistent with
relativity, and no true paradoxes can be found. However, several
apparent paradoxes have been extensively discussed and these
discussions enriched our understanding of the electromagnetic theory.
Some of these controversial topics are: hidden momentum \cite{ShJa},
Feynman's disk \cite{FD}, Trouton-Noble Experiment\cite{TN}, and the 4/3
factor for the self energy of an electron.\cite{43} Here we present
another situation which, analyzed in a naive way, leads to paradoxical
conclusions. The paradoxes are relevant to recent discussions of
covariance in the electromagnetic
theory\cite{Je99,Hn98,Co00,Hn00,Go00}.

Let us start with  the paradox which is simplest to present; its
resolution will be shown at the end of the paper.

\vskip 0.2cm
\noindent
{\bf Paradox I}
 
There are  two particles of charge $q$ initially separated by distance
$l$.  We consider two ways to bring the particles, initially and
finally at rest, to a shorter
distance $l-x$, see Fig.~1.

(i)  We move   one particle the distance $x$ toward the other particle. The work required for this is:
\begin{equation}
\label{W}
W^{\rm i} = U_{NEW} - U_{OLD} = {q^2 \over {l-x}} -   {q^2 \over {l}} .
\end{equation}

(ii) We move both particles toward each other for the distance $x/2$.
We do it simultaneously and fast enough such that the motion of each
particle ends before the signal about this motion can reach the
location of the other particle. In this case, the work should be the
sum of the amounts of work performed by external forces exerted on the
two particles calculated as if the other particle has not moved:
\begin{equation}
W^{\rm ii} = W_1 + W_2 =2 \left( {q^2 \over {l-{x\over 2}}} -   {q^2 \over {l}}\right) .
\label{W'}
\end{equation}
\vskip .4cm

\begin{center} \leavevmode \epsfbox{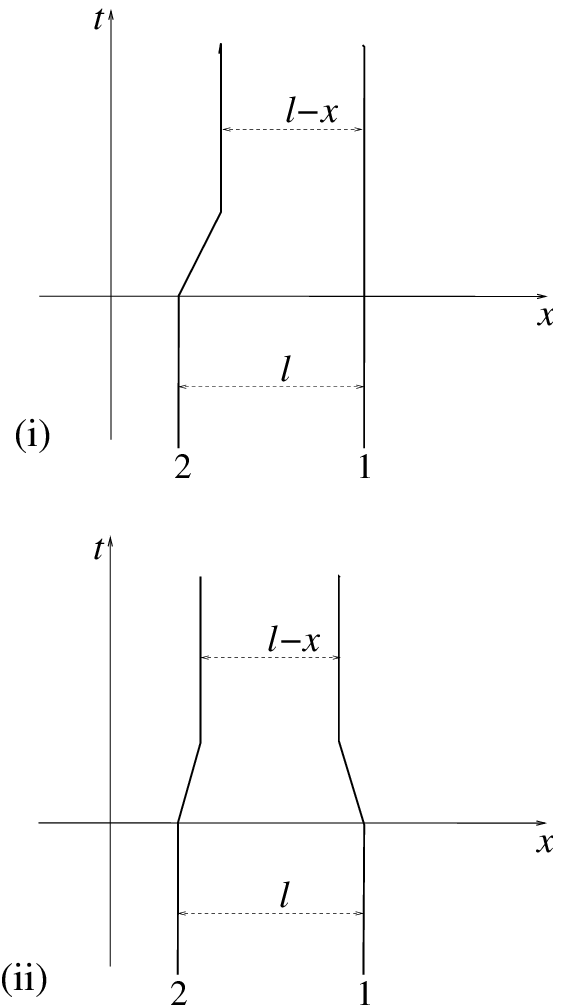} \end{center}

\noindent 
{\small {\bf Fig.~1.} Space-time diagram of the motion in the two
  processes: (i) one particle moves, (ii) two particles move.}
\vskip .4cm
After the procedure is ended, we obtain the same situation in both
cases, but we applied less work when we moved both particles: $W^{\rm
  ii}<W^{\rm i}$.

We can get the energy equal to the work $W^{\rm i}$ back from the system when
we reverse the process (i), moving one of the charges to the original
distance~$l$.

\noindent
 We can repeat the cycle of process (ii) followed by
reversed process (i) gaining every time the amount of energy:
\begin{equation}
\label{gain}
W^{\rm i}- W^{\rm ii}  \approx {q^2 x^2 \over {2l^3}}.
\end{equation}

Of course, there must be an error in the above argument. We have not
taken all relevant effects into account. However, before explaining
this paradox we present and resolve a few other apparent paradoxes
demonstrating the relevance of various effects. In Section II we
present and in Section III resolve a paradox based upon part of the
process (ii) described above. Instead of starting with particles at
rest, accelerating, moving with some velocity, and then bringing the
particles to rest, we start with the two particles moving with
constant velocity and coming to rest at different times.  In Section
IV we further analyze the setup of Paradox II obtaining and resolving
another paradox. Section V explains the same point using an example of a
large number of charged particles. In Section VI we consider remaining
part of the processes involved in Paradox I: accelerating a pair
of particles from rest.  In Section VII we consider acceleration of
particles moving in parallel. This setup leads to a paradox the
resolution of which is due to yet another surprising effect. In
Section VIII we return to the analysis of Paradox I and resolve it in
a quantitative way.

\section{ PARADOX II: CONSERVATION OF ENERGY FOR TWO STOPPING PARTICLES}
\label{para1}

Consider two particles of charge $q$ and  mass $m$ located on the $x$ axis at
the separation $l$ and moving in the $x$ direction with a constant velocity
$v$. At time $t_1=0$ we stop the first particle and at time $t_2 =t$ we
stop the second particle: see  Fig.~2. The time $t$ is small  enough such
that  signals about the change of the velocity of one particle cannot reach the
location of the other, while it is still moving:
\begin{equation}
\label{tbound}
  {-l\over {c - v}} < t < {l\over {c + v}}. 
\end{equation}

\noindent
We also require that $\tau$, the time  of ``stopping''
a particle, is small: $\tau \ll |t|$. 

Let us consider conservation of energy for this process. The initial
energy should be equal to the final energy plus the work of the forces
which the particles exert on external systems:
\begin{equation}
\label{conserv}
E_{in} = E_{fin} + W_1 +W_2 +\tilde W,  
\end{equation}
where $W_1$ and $W_2$ are the works due to the forces the particles
exert during the process of stopping; $\tilde W$ is the work performed
by the particle moving with velocity $v$ during the time that the
other particle is at rest (for $t>0$, the work $\tilde W$ is performed
by particle 2 and for $t<0$, by particle 1). Of course, no work is
performed when both particles are at rest, and, also, the net work
vanishes during the time when both particles are moving with velocity
$v$.

For making relativistic analysis more convenient, we include the
rest mass energy. Then, the final energy of the system is 
\begin{equation}
\label{Efin}
E_{fin}= 2mc^2 + {q^2 \over {l - x}} ,
\end{equation}
where $x$ is the change   in the distance between the charges:
\begin{equation}
\label{xvt}
x = vt .
\end{equation}
The distance might decrease or increase (negative $t$ and $x$)
depending on which particle stopped first.

When a particle moves with constant velocity, the total force exerted
on it is zero. Therefore, the force it exerts on an external system is
equal to the electromagnetic force the other particle exerts on it.
Since, in the laboratory frame, the distance between the particles is $l$,
in the Lorentz frame at which the  charges are at rest, the distance
between them is $\gamma l$ (where $\gamma~\equiv~1/\sqrt{1-v^2/c^2}$). The Lorentz transformation for the force in the $x$
direction between the rest frame and the laboratory frame is $F_x=
F'_x$ and, therefore, the forces the particles exert on the external
systems are:
\begin{equation}
\label{force}
  {F_1}_x = - {F_2}_x = {q^2 \over {(\gamma l)^2}}.
\end{equation}
 Thus, the work $\tilde W$ is:
\begin{equation}
\label{tilda}
\tilde W = - {{q^2 x}\over {(\gamma l)^2}} .
\end{equation}

This formula is correct both for $x > 0$, when the work is done by
particle 2,  and for $x < 0$, when the work is done by
particle 1.
Thus, the equation of conservation of  energy  (\ref{conserv}) becomes 
\begin{center} \leavevmode \epsfbox{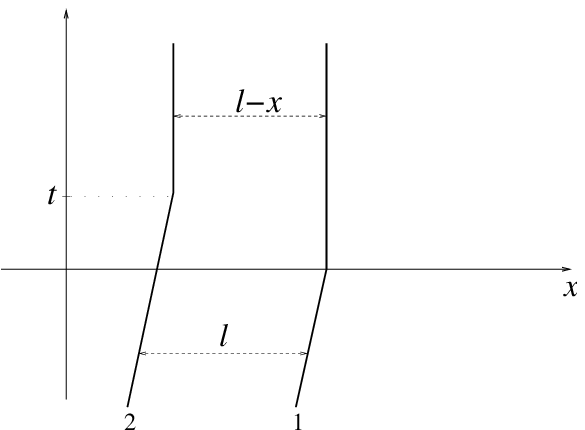} \end{center}
\noindent 
{\small {\bf Fig.~2.} Space-time diagram of the motion of the two particles.}

\break

\noindent 
\begin{equation}
\label{conserv1}
 E_{in} =  2 mc^2 + {q^2 \over {l - x}}+  W_1 +W_2     - {{q^2 x}\over {(\gamma l)^2}}. 
\end{equation}
 The initial energy $E_{in}$ obviously does not depend on $x$. Due to
causality argument, the works $ W_1$ and $W_2$ do not depend on $x$ either.
The equation of energy has only two terms depending
on $x$. Therefore,  Eq.  (\ref{conserv1})
represents a paradox: it must be true for all $x$ in the
interval $[{-lv\over {c - v}},{lv\over {c + v}}]$ (corresponding to
(\ref{tbound})), but it cannot, since it has only two terms depending
on $x$ which do not balance each other.

\section{RESOLUTION OF PARADOX II: INTERFERENCE OF RADIATION}
\label{PARII}

In Paradox II, according to our naive calculation, the
final energy together with the obtained work had two terms depending
on $x$, the change in the distance between the particles, which do not
sum up to a constant. There is cancellation of the $x$ dependence in
the first order of the parameter ${x\over l}~\sim~{v \over c}$,
therefore, the leading term  of the unbalanced $x$ dependent term  is
$ {{q^2 x^2}\over {l^3}}$. Indeed,
\begin{equation}
\label{2term}
 {q^2 \over {l - x}}   - {{q^2 x}\over {(\gamma l)^2}}
= {q^2\over l} +  {{q^2 x^2}\over {l^3}} + ...
\end{equation}
 The process of stopping cannot be arbitrarily slow since
it had to be finished before the signal about stopping the other particle
can arrive. Moreover, we choose $\tau \ll |t|$. Therefore, we
should expect a significant   contribution due to {\em
  radiation} which  we have not taken
into account.
 The charges accelerate during the process of stopping and
the magnitude of the acceleration is $a = {v \over \tau}$~.  According
to the Larmor formula, the total energy radiated by a single charge
during the whole process of stopping is
\begin{equation}
\label{rad}
 R_1=R_2= {2 \over 3} {{q^2 a^2}  \over c^3}  \tau = {2 \over 3}
 {{q^2 v^2}  \over { c^3\tau}}. 
\end{equation}
The $x$ dependent term which we have to balance is much smaller than $R_1$ and $R_2$: 
\begin{equation}
\label{2term<}
  {{q^2 x^2}\over {l^3}} = {{q^2 (vt)^2}\over {l^3}} \lesssim {{q^2
      v^2}\over {c^2 l}} \ll  {{q^2 v^2}\over {c^3 \tau}}.
\end{equation} 
However, everything that happens at the close vicinity of the charges
cannot depend on $x$, and, in particular,  the radiation which each
charge emits does not depend on $x$, so how can the radiation energy
balance the $x$ dependent terms in the equation of conservation of
energy? The effect is due to the {\it interference of radiation}.  
The total radiated energy is 
\begin{equation}
\label{Rtot}
  R_{tot}= R_1+R_2+R_{int} .
\end{equation} 
The interference term $R_{int}$ depends on $x$ and restores the
balance. Now we will show this in detail.

The radiation of the stopping charge propagates inside a spherical
shell of width $c \tau$ and the energy flux {\bf S} is given by:
\cite{Grif}
\begin{equation}
\label{radS}
 {\rm \bf S}   =  {{q^2 a^2 \sin^2 \theta}  \over {4 \pi c^3 r^2}}
 {\bf \hat r},
\end{equation}
where $r$ is the radius of the shell and  $\theta$ specifies the direction
relative to the $x$ axis in our setup. Since we have two accelerated
charges, the radiation field due to the two charges will interfere in the
region of the overlap, see Fig.~3.
 The complete overlap will take place for the angle
$\theta$ defined by:
\begin{equation}
\label{theta}
\sin (\theta -{\pi\over2}) ={{c t} \over{ l-vt}}= {{c x} \over{v(l-x)}}.
\end{equation}  
Since the width of the shells is $c \tau $, the overlap will be
nullified beyond the deviation $\delta \theta$ from the angle $\theta$ given by
\begin{equation}
\label{thetaD}
 \delta \theta ={{c  \tau} \over {(l-x) \sin \theta}} ~,
\end{equation}  
which is obtained from  
\begin{equation}
\label{thetaDer}
c \tau =\delta \left[(l-x)\sin (\theta -{\pi\over2})\right] =  [(l-x)\sin \theta ]\ ~\delta \theta.
\end{equation}  
Due to the interference,  the total
energy radiated in the direction of the overlap is twice as much as if
the two charges were radiating separately. At the interval $[(\theta - \delta \theta),(\theta + \delta \theta)]$
the overlap increases and then decreases linearly.
 Therefore, the interference term of the radiation  energy is

\begin{equation}
\label{rad-int}
 R_{int} =
 2 {{q^2 a^2 \sin^2 \theta}  \over {4
     \pi c^3 r^2}} 2\pi r \sin \theta \, r
  \tau \int_{- \delta \theta}^{\delta \theta} { {\delta \theta - |\phi |}\over \delta \theta} d\phi =
 {{q^2 v^2 \sin^2 \theta}  \over { c^2 (l-x)}} .
\end{equation}

\begin{center} \leavevmode \epsfbox{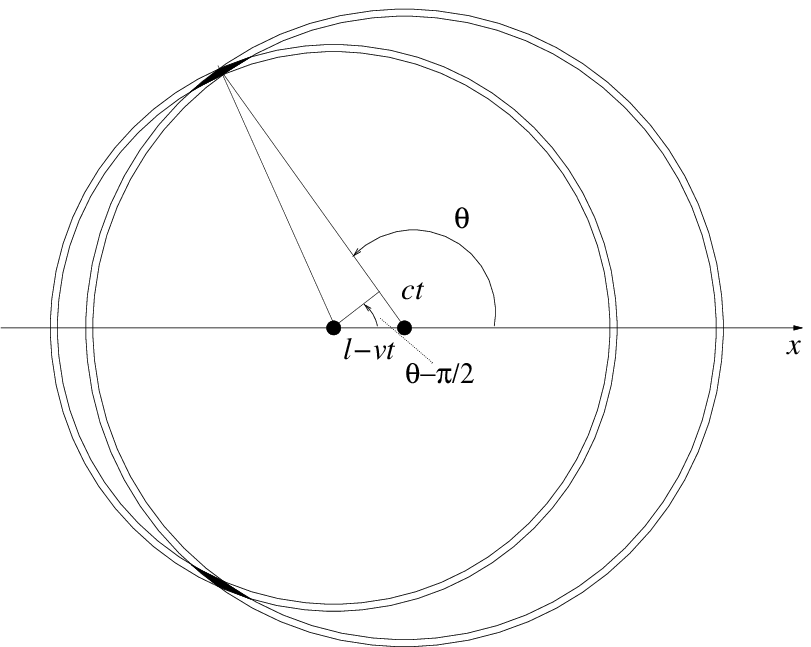} \end{center}

\noindent 
{\small {\bf Fig.~3.} Electromagnetic radiation of the two stopping particles. The area of constructive interference of the radiation field is painted in black.}

\noindent
After expressing $\sin^2 \theta = 1-
\sin^2 (\theta -{\pi\over 2})$, using (\ref{theta}), and making an approximation up  to the lowest order in  the parameter ${x\over l} \approx {v \over c}$,
we obtain:
\begin{equation}
  \label{Eradint}
 R_{int}=
 {{q^2 v^2 }  \over { c^2 (l-x)}} - 
 {{q^2 x^2}\over {(l-x)^3}} \approx    {{q^2 v^2 }  \over { c^2 l}} - 
 {{q^2 x^2}\over {l^3}} .
\end{equation}
 Thus, we have shown that (at least up to a
second order in the parameter ${v \over c}$, the precision to which we
made our calculations) the equation of conservation of energy which
takes into account the electromagnetic radiated energy does not have
$x$ dependence. The corrected equation of conservation of energy  (which replaces Eq.
5),
\begin{equation}
\label{conserv+rad}
E_{in} = E_{fin} + W_1 +W_2 +\tilde W + R_{tot},  
\end{equation}
leads, after the approximation, to the expression which does not exhibit $x$ dependence:
\begin{equation}
\label{conserv3}
 E_{in} =  2 mc^2 + 
 W_1 +W_2    +R_1 +R_2 +{{q^2 }  \over {  l}}\left( 1 + {{ v^2 }  \over { c^2}}\right).
\end{equation}
Thus we have 
 resolved  Paradox II with regard to the unbalanced $x$ dependence. But can we
 show that the LHS and the RHS of the equation of conservation of
 energy (\ref{conserv3}) are equal? We  will analyze this  in the
 next section.

The paradox of non-conservation of energy of
the system of two charged particles when radiation is neglected has
been considered in another example.\cite{scat73} The calculation of
the scattering  on the basis of
the Coulomb forces yields an energy of the charged
particles after the collision that is larger than the initial energy.
The advantage of the scattering example is that no external forces
have to be taken into account. However,  the resolution of this
paradox by taking into account radiation has been shown only
qualitatively.\cite{scat76}

\section{LORENTZ TRANSFORMATION FOR ELECTROMAGNETIC ENERGY }
\label{PARIII}

Without knowing the mass and without knowing the local works $W_1$ and
$W_2$ it seems that we cannot test (\ref{conserv3}). However, we can
test the consistency of this equation of conservation of energy with
single-particle conservation of energy equations.  We can write down
the conservation of energy for each particle assuming that it performs
exactly the same motion (stopping from velocity $v$ during the time
$\tau$) in case the other particle is not present. We can argue that
the local works $W_i$ and the emitted radiations $R_i$ are the same as
in the original example. 

This argument is not as strong as the causality argument, i.e. the
independence of $x$ of the values of these variables, but it seems
that since we can take the time $\tau$ of the local processes very
small, the effect of the external field can be neglected. In order to
reduce any doubt that this is a valid approach, we will consider, in
the next section, a similar situation with a large number of charged
particles. It exhibits the same problem, but we will not need this
assumption.

Our approach to finding the initial energy is finding the total energy of
the two charges in their rest frame $E_0$ and multiplying it by the factor
$\gamma$. 
In the rest frame of the moving particles, the distance
between them is $\gamma l$. Therefore, the total initial energy of the
particles is:
\begin{equation}
\label{Ein}
E_{in}= \gamma \left( 2mc^2  + {q^2 \over {\gamma l}}\right) .  
\end{equation}
Thus, the equation of conservation of energy  (\ref{conserv3}) becomes 
\begin{equation}
\label{conserv4}
 2\gamma  mc^2+ {{q^2 }\over {l}} =  2 mc^2 + 
 W_1 +W_2    +R_1 +R_2 + {{q^2 }  \over {  l}}\left( 1 + {{ v^2 }  \over { c^2}}\right).
\end{equation}

We can
write two (identical) single-particle equations of conservation of energy:      
\begin{eqnarray}
\label{conserv1p}
 \gamma  mc^2 =   mc^2 + 
 W_1     +R_1 , \nonumber \\
 \gamma  mc^2 =   mc^2 + 
 W_2     +R_2 .
\end{eqnarray}
But when we subtract these equations from the two-particle
conservation of energy equation (\ref{conserv4}) we see that there is
inconsistency: the term ${{q^2 v^2}\over {l c^2}}$ is unbalanced! The
 inconsistency does not follow from the approximations we made
in deriving  (\ref{conserv3}).  The algebraic approximations can be made irrelevant if we
consider simultaneous stopping of the charges  corresponding to
$x=0$, in which case 
 (\ref{conserv3}) follows  without
 approximation. We have reached another paradox. There must be another
 error in our analysis. (In fact, this paradox will appear also for
 large $|x|$ when there is no interference of radiation; such a case
 will be considered in the next section.)

The paradox arises from the error  which we made in Eq.
(\ref{Ein}). We have claimed that if the total energy of a system of
charges in their rest frame is $E_0$, then its energy in the Lorentz frame
in which the system moves with velocity $v$ is
\begin{equation}
E = \gamma E_0 .
 \label{gama}
\end{equation}
Equation (\ref{gama}) is, of course, correct when the system is an
elementary particle. It is also true when the system is a finite
stationary isolated body.   But it is, in general, not true for
a composite system with external forces such as the system of charged
particles we consider here. 

In order to obtain the correct transformation of the electromagnetic
energy from the rest frame to the moving frame, we consider two
charges connected by a rigid rod. The energy of the whole system,
charges and the rod, does transform according to (\ref{gama}).
Therefore, the anomalous term of the transformation of the
electromagnetic energy equals to the negative of the anomalous part of
the mechanical energy of the rod. The latter is easier to calculate
and we will do it now.

In order to calculate the expression for transformation of energy of
the rod, we  express it as a volume integral of the energy density $u$:
\begin{equation}
  \label{enden}
E_{in} = \int u dv  ,
\end{equation}
 and  use  the Lorentz transformation for the energy density,
 the 00 component of the
energy-stress tensor:
\begin{equation}
u = \gamma ^2 \left(u' + {v\over c^2} S'_x  -{v^2\over c^2} \sigma'_{xx}\right),
 \label{uu}
\end{equation}
where ${\bf S}$ is the energy flux and   $\sigma$ is the stress tensor. 
The transformation of the energy due to the the first term leads to the
usual expression  (\ref{gama}): the energy density is multiplied by
$\gamma^2$, but due to the Lorentz contraction the volume is multiplied by
the factor $\gamma^{-1}$. The second term does not contribute since
the energy flux in the rest frame vanishes. Therefore,  the last term
contributes the anomalous term.
 The tension in the
rod prevents the charges, separated by the distance $\gamma l$, from moving, therefore it equals
${q^2} \over {\gamma^2 l^2}$. Thus, in the rest frame of the rod, the  stress
  tensor component is:
\begin{equation}
  \label{tens} 
  \sigma'_{xx} = {q^2\over {s \gamma ^2l^2}}~,
\end{equation}
where $s$ is the cross-section of the rod.
  Therefore, the  contribution to the energy in
the laboratory frame due to the tension of the rod
 is:
\begin{equation}
  \label{cont}
-{{\gamma^2 v^2}\over c^2}\int \sigma'_{xx} dv = -{{\gamma ^2 v^2}\over c^2}  \sigma'_{xx}
ls = - {{v^2 q^2}\over {c^2 l}}  .
\end{equation}
Thus, the correct expression for the initial energy of the
electromagnetic field of the two charges  (including the anomalous term
of the transformation of the electromagnetic energy which equals to
the negative of that of the mechanical energy) is:
\begin{equation}
\label{Einmod}
E_{in}= \gamma\left[ 2mc^2  + {q^2 \over {\gamma l}} \left(1 +  {{ v^2}\over {  c^2}}\right) \right] .  
\end{equation}
The correction we found for the initial energy of the system of two
charges equals exactly the interference term of the radiation energy thus
restoring the equation of conservation of energy. 

\section{ CONSERVATION OF ENERGY FOR $N$ STOPPING PARTICLES}
\label{Npar}

In this section we can strengthen the arguments of the previous
section by considering a large number of charged particles in a row.
In this case, we do not subtract single-particle equations of
conservations of energy from the $N$-particle equation of conservation
of energy. Thus, we do not need the assumption of the previous section
that the terms of single-particle equations are equal to the
corresponding terms of the $N$-particle equation. However, since this
section does not describe conceptually new effects, it can be omitted
on the first reading.

Consider a chain of
a large number $N$ of identical particles of charge $q$ and mass $m$ separated by  the distance $l$ one
from the other,  all moving with velocity $v$ on
the $x$ axis. The first particle stops at $t_1=0$ during a short time $\tau$. The
second particle stops in the same manner at $t_2=t$, the latest possible moment
such that the information about stopping of the first particle cannot
reach it. This corresponds to
\begin{equation}
x = vt = {vl\over {c+v}} .
\label{xmax}
\end{equation} 
The third particle stops in the same way at time $t_3 = 2t$ just
before the information about the stopping of the first two should
arrive, the particle $n$ stops at time $t_n = (n-1)t$, etc.  until the
stopping of the last particle. This  is
illustrated in Fig.~4.

One difference from the case of two particles is that, due to
the particular choice of $x$, we have no interference of radiation from
different charges. The overlap of the radiation fields takes place
only on the $x$ axis where the intensity is zero. Another  
 change from the case of two particles to the case of $N$ is
reflected in the calculations of the initial and final potential energies,
the work $\tilde W$ (the work made by particles during motion with constant velocity), and the anomalous energy transformation term: in
all these terms appear  the factor $\eta$:
\begin{equation}
\label{eta}
\eta= \sum_{n=1}^{N-1} \sum_{i=1}^n {1\over i}. 
\end{equation}

The appearance of the factor $\eta$ is obvious for potential energy,
since ${q^2\over {\gamma l}}\sum_{i=1}^n {1\over i}$ is the energy of bringing
the particle number $n+1$ when there are $n$ particles in the row. Let
us show that the same factor appears for $\tilde W$.  When all
particles move with velocity $v$, no net work is done on the system
and, of course, no net work is done when all particles are at rest.
Consider forces between particle $n$ and all particles $i$ to the
right of it, (i.e., $i<n$). The distance the particle $n$ moves while
$i$ is at rest is $(n-i)x$, and the force between them is ${{q^2}
  \over {(\gamma l (n-i))^2}}$. Therefore, the contribution to $\tilde
W$ due to the interaction between particle $n$ and particle $i$ is
\begin{center} \leavevmode \epsfbox{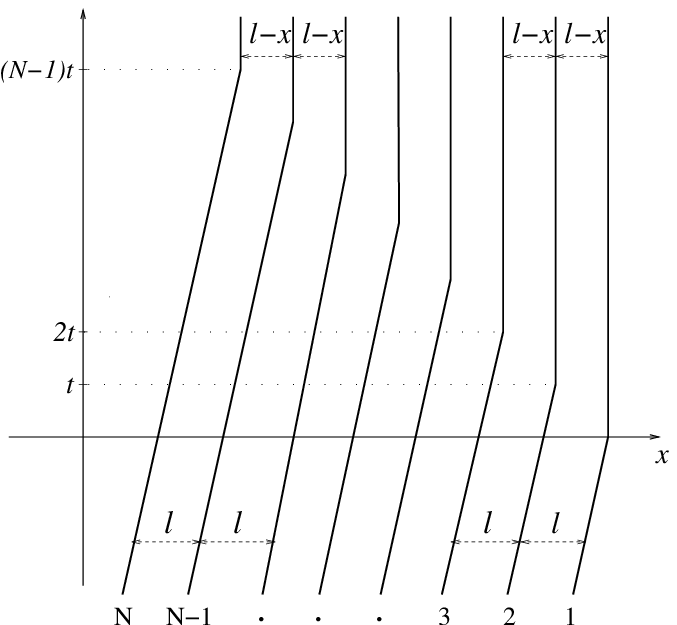} \end{center}
\noindent
{\small {\bf Fig.~4.} Space-time diagram of the motion of $N$ particles.}

\begin{equation}
  \label{con}
{{q^2 x} \over {(\gamma l)^2(n-i)}}.
 \end{equation}
 We obtain that the contribution
to the potential energy due to the forces between particle $n$ and all
particles $i$ such that $i<n$ is:
\begin{equation}
  \label{contri}
\sum_{i=1}^{n-1} {{q^2 x} \over {(\gamma l)^2(n-i)}} ={{q^2 x} \over {(\gamma l)^2}} \sum_{i=1}^{n-1} {1\over i}.
 \end{equation}
The sum of the works made by all particles, starting from the moment when the first particle stops, is:
\begin{equation}
  \label{tild}
\tilde W ={{q^2 x} \over {(\gamma l)^2}} \sum_{n=2}^{N}
  \sum_{i=1}^{n-1} {1\over i} =\eta {{q^2 x} \over {(\gamma l)^2}}.
\end{equation}

Instead of performing a direct calculation of the anomalous
transformation term (deviation from (\ref{gama})), we can make the following observation. The
anomalous term can be found by calculation of the contribution due to  the tension of the rod
(compare with (\ref{cont})):
\begin{equation}
  \label{contN}
-{{\gamma^2 v^2}\over c^2}\int \sigma'_{xx} dv =
 -{{\gamma ^2 v^2}\over c^2}\sum_{n=1}^{N-1}  (\sigma'_{xx})_n ls =
 - {\gamma ^2 v^2\over c^2} \sum_{n=1}^{N-1}  T_n l .
\end{equation}

Consider a rod with $N$ charges separated by distance $l$ at
rest. We will show that 
 $\sum_{n=1}^{ N-1} T_n l$ is equal to  the  potential energy of the
 charges. Since the latter is multiplied by $\eta$ in the transition
 from 2 to $N$, the former is multiplied by $\eta$ too.
Potential energy is  equal to  the  work which
electromagnetic forces will do in the process of uniform extension of
the length of the rod from $(N-1)l$ to a very large length, which, in turn,
is equal to the negative of the mechanical work made by the tension
forces of the rod. When the separation between the charges is $\tilde
l$, the tension in the parts of the rod can be expressed as
\begin{equation}
   T_n (\tilde l)= T_n (l) { l^2\over \tilde l^2}~,
\end{equation}
because this tension compensates the Coulomb force proportional to $\tilde l^{-2}$. Therefore, the work (equal to the potential energy) can, indeed, be expressed as
 \begin{equation}
  \label{work}
\sum_{n=1}^{N-1}
 \int_l^{\infty} T_n (\tilde l) d\tilde l =\sum_{n=1}^{N-1}
 \int_l^{\infty} 
T_n (l) { l^2\over \tilde l^2} d\tilde l =\sum_{n=1}^{N-1} T_n l .
\end{equation} 

Now we can write the equation of conservation of energy for $N$
charges including the anomalous transformation term 
(the LHS is the modification of (\ref{Einmod}) and the RHS is the modification of the RHS of (\ref{conserv1})): 
\begin{eqnarray}
\label{conserv2}
\nonumber
 N\gamma  mc^2+ \eta {{q^2 }\over {l}}\left( 1 + {{ v^2 }  \over { c^2}}\right) =~~~~~~~~~~~~~~~~~~~~~~~~~~~~
\\ N  mc^2 +\eta {q^2 \over {l - x}}+
  W_1 +W_2 + ... + W_N - \eta {{q^2 x}\over {(\gamma l)^2}}. 
\end{eqnarray}

In order to estimate $\eta$ for large $N$ we can replace the sums by
the integrals (and $N-1$ by $N$):
\begin{equation}
  \label{eta-in}
 \eta =  \sum_{n=1}^{N-1} \sum_{i=1}^n {1\over i}    \approx
 \int_{1}^{N}\left ( \int_{1}^z {{dy}\over y}  \right ) ~ dz \approx \int_{1}^{N} \ln
 z dz \approx  N\ln N  .
\end{equation}
(We omitted $N$, since it is negligible relative to $N \ln N$).

Since for large $N$ we have $\eta \gg N$, all terms in
(\ref{conserv2}) which do not have the factor $\eta$ can be neglected.
These include all terms $W_n$, the amount of work obtained in the
process of stopping particle $n$. Indeed, for small enough $\tau$,
these terms will not have strong dependence on $N$ and, therefore,
their sum will be approximately proportional to $N$.  After
substitution of (\ref{xmax}), our choice of $x$, we can see that the
equation of energy with the anomalous transformation term is balanced
in the leading $\eta$ proportional terms (and it is not balanced if we
use  the transformation of energy (\ref{gama})).

\section{ ACCELERATING PARTICLES FROM REST }
\label{acc}

In order to prepare ourselves for the analysis of Paradox I presented
in the Introduction, let us consider a step which has not been
discussed yet: the acceleration of the two charged particles from rest to
velocity $v$. Of course, in the frame of reference moving with
velocity $v$, this is deceleration and stopping of particles moving with equal velocities,  the  process  we analyzed
above. However, the transformation from one
frame to another might be a difficult task and,  as in many other examples
\cite{BK,Nami}, an analysis in a different Lorentz frame allows seeing new
physical phenomena; in our case it provides yet another possibility to
make an error leading to a paradox. 

We accelerate the
particles in the same way as in the process of
simultaneous deceleration discussed in the preceding paragraph, i.e.,
we perform simultaneous acceleration from rest to velocity $v$ during
 time $\tau$. The radiated energy  during the acceleration, then,
 should be the same as in the process of stopping, see Eq. (\ref{Eradint}). For simultaneous
 acceleration $x=0$, so the interference term is
\begin{equation}
  \label{Era1}
 R_{int} =   {{q^2 v^2 }  \over { c^2 l}} .
\end{equation}
Initial energy of the particles is
\begin{equation}
E_{in} =  2mc^2 + {{q^2 }\over {l}} ,
 \label{EinN}
\end{equation}
Since the final state of the particles (motion with velocity $v$ and
separation $l$) is identical to the initial state of the particles in
the previous example, the expression for the final energy $E_{fin}$ is
given by the RHS of  (\ref{Einmod}).

Again, for each particle we can write the equation of conservation of energy for the process when the other particle is not present:
\begin{eqnarray}
\label{conserv1p2}
 mc^2 =  \gamma  mc^2 + 
 W_1     +R_1 ,\nonumber \\
   mc^2 = \gamma  mc^2 + 
 W_2     +R_2 ,
\end{eqnarray}
where $W_1$, $W_2$ are defined as work performed by the particles and
are negative in this case.
 
 Since the charges start to move together, it seems that the net work
 is done only during the acceleration period.
 Therefore, the
equation of conservation of energy is:
\begin{equation}
 E_{in} = E_{fin}+
    W_1 +W_2 +R_1 +R_2 + R_{int} .
 \label{Conserv4}
\end{equation}
Substituting all the terms  and subtracting the single-particle
equations  we obtain: 
\begin{equation}
  \label{contr}
 {{q^2 }\over   l }=   
  {{ q^2} \over
   l}\left(1 + {v^2\over c^2}\right) + {{ q^2 v^2}\over   {l c^2}}.
 \end{equation}
The final energy is larger than the initial energy: contradiction!

The error we made here is more transparent. It appears in the sentence
stating that the only net work of the charges is done during the
acceleration period. It is true that in a stationary case (particles
keep their motion all the time) the net work of charges moving with
constant velocity vanishes. However, at the beginning of the motion,
the fields at the vicinity of the charges are different from the
Coulomb field of the stationary motion: each particle feels the {\it
  static} field of the other particle (i.e., as if it has not moved)
until the signal from the motion of the other particle can arrive.

Let us  calculate the contribution to the work due to the forces
between the particles.
 Particle 2 moves in the
static field of particle 1 during the time $t={{l}\over {c+v}}$
after which it feels the stationary field of particle $1$ which is ${q
  \over {\gamma ^2 l^2}}$.
Similarly, particle 1 moves in the static field of particle 2
during the time $t'={{l}\over {c-v}}$, after which it feels the
stationary field of particle 2 of the same strength but in the opposite
direction. After time $t'$, there is no contribution to the net work
due to the forces between the two particles. Until this time, particle
1 covers distance $x'=vt'$ in the static field of particle 2.
Therefore, the contribution to the work from particle 1 is:
\begin{equation}
 \label{icont}
 { {q^2 }\over   {l} }- {{q^2 }\over   {l+x'} }
={ {vq^2 }\over   {cl} } .
 \end{equation}
 The work performed by particle 2 until time $t'$ has two parts.
 Until time $t$, it is:
\begin{equation}
 \label{ncont}
 { {q^2 }\over   {l} }- {{q^2 }\over   {l-x }}=
 -{ {vq^2 }\over   {cl} } .
 \end{equation}
Between time $t$ and $t'$ it feels a constant field so the contribution
to the work is:
\begin{equation}
 \label{ncontt1}
-{q^2 \over {\gamma^2 l^2}}v(t'-t)= -{ {2 v^2 q^2 }\over   {c^2 l} } .
 \end{equation}
The contributions (\ref{icont}) and  (\ref{ncont}) cancel each other, so
the net contribution is given by  (\ref{ncontt1}). The net work performed
by the particles  during the time of motion with constant velocity is:
\begin{equation}
 \label{ncontt}
\tilde W =  -{ {2 v^2 q^2 }\over
  {c^2 l} } .
 \end{equation}
This restores the balance in the equation of conservation of energy
canceling the unbalanced terms in (\ref{contr}).

We have resolved the contradiction by  taking into account the work
performed by charged particles
during the transition period from static field to stationary field
which lasts ${{l}\over {c-v}}$.
 But what happens if we  skip the
intermediate stage? We accelerate the particles to velocity $v$ and
shortly after (before the particles finished performing the work
(\ref{ncontt}))  stop them in a similar manner. What is the source of
the radiation energy (\ref{Era1})  in this case? In fact, since we
have two processes with the acceleration, it seems that we are missing 
even more, twice
this amount!

No, this is another error. We need not look for the source of the
radiation energy, because the radiation field due to the acceleration
and the radiation field due to stopping interfere destructively. (Note
a more bizarre example of a destructive interference of radiation
field from a moving body.\cite{nonrad}) We are not going to analyze the destructive interference in a quantitative way in this case. The same  effect yields the resolution
of Paradox I which is demonstrated in a   quantitative way in   Section
\ref{PARI}.   Before this, in the next section, we present   a 
paradox arising from yet another subtle  effect.

\section{ ACCELERATING PARTICLES MOVING IN PARALLEL  }
\label{accc}

Let us consider acceleration of two charged particles lined up in the $y$ axis
instead of the $x$ axis.  The particles accelerate simultaneously from rest to velocity $v$ in the $x$ direction. The expression for the initial energy is
again  (\ref{EinN}).
 The final energy, however, is different:
\begin{equation}
\label{Efin2}
E_{fin}= \gamma\left( 2mc^2  + {q^2 \over { l}}  \right) .  
\end{equation}
Indeed, in the rest frame of the moving particles the distance between
them is $l$. In this case the electromagnetic energy is transformed in
the usual way (\ref{gama}) because the energy of the composite system
of charges and the rod is transformed according to  (\ref{gama})
and the energy of the rod with the tension in the $y$ direction is  transformed according
to (\ref{gama}). The anomalous behavior of the rod in the previous
case followed from the presence of $\sigma_{xx}$ component of the
stress tensor which vanishes in the vertical configuration.

The interference term of the radiation energy is modified too. In
the $x$ axis configuration, the interference is in the $y-z$ plane
and it always has the angle $\pi \over 2$ relative to the direction of the
acceleration. In the  $y$ axis configuration, the interference is in the $x-z$ plane
 with varying angle $\theta$ relative to the direction of the
acceleration. Therefore, the interference is not always  in the
direction of the maximal intensity. The intensity is proportional to
$\sin^2\theta$, see (\ref{radS}), and, therefore, averaging on $\theta$
 reduces the interference term relative to the $x$
configuration (\ref{Era1}) by the factor of 2:
\begin{equation}
  \label{Era11}
 R_{int} =   {{q^2 v^2 }  \over {2 c^2 l}} .
\end{equation}

Although during the stationary motion the charges do not exert forces
in the direction of motion, in the transition period, when the charges
move in the static field, there is a small component of the force in
the direction of motion. The particles move in the static field
during the time  $t$ which fulfills
\begin{equation}
  ct=\sqrt{ l^2 +t^2 v^2}.
\label{time1}
\end{equation}
Therefore $ ct=\gamma l$.  The total work each particle performs is
\begin{equation}
  \tilde W_1= \tilde W_2={ q^2 \over l} - { q^2 \over{\gamma l}}.
\label{time11}
\end{equation}
The single-particle equations of conservation of energy, of course, remain the same. Thus, putting together all terms of the conservation of energy equation 
\begin{equation}
 E_{in} = E_{fin}+
    W_1 +W_2+
   \tilde  W_1 + \tilde W_2  +R_1 +R_2 + R_{int},
 \label{Conserv44}
\end{equation}
and subtracting single particle equations (\ref{conserv1p2}), we obtain:
\begin{equation}
 \label{contr5}
 {{q^2 }\over   l }=  {\gamma q^2 \over  l}  +  {2 q^2\over l} \left(1
   -{1\over \gamma}\right)+ {{q^2 v^2 }  \over {2 c^2 l}} .
   \end{equation}
Contradiction again: The final energy is larger than the initial
energy. Calculating up to the second order in the parameter $v^2 \over
c^2$ we see  three contributions in the units of ${ {v^2q^2 }  \over {
  c^2 l}}$. The increase in  the potential energy contributes $1\over 2$, the
work of the static fields during the transition period contributes
$2\cdot {1\over 2} =1$, and the interference of radiation contributes
$1\over 2$.  All terms together contribute  ${ {2v^2q^2 }  \over {
  c^2 l}}$. 

The effect we missed here is, probably, the most  subtle one. We have not
taken into account the work of the radiation field.  The electric field at the point
${\bf r}$ (relative to the charge), due to the
radiation of the charge $q$ moving with 
 acceleration ${\bf \it a}$,
is: \cite{Grif}
\begin{equation}
  \label{Efrad}
  {\bf E} = {q\over {c^2 r}} {\bf \hat r} \times  ({\bf \hat r}
 \times {\bf \it  a}).
\end{equation}
For our configuration, the field is: 
\begin{equation}
  \label{Efrad1}
  {\bf E} = -{{q a}\over {c^2 l}} { \hat x} .
\end{equation}
This field exerts force during the time $\tau$ during which the
particle moves distance $\tau v$. Taking into account that $a =
{v\over \tau}$ we obtain that the force  of the radiation field
changes the energy of each
particle by 
\begin{equation}
W_{rad}= - {{q^2 v^2 }  \over {
  c^2 l}}.
\end{equation}
Both particles lose their energy in this way and, therefore, we lose two
units of ${{v^2 q^2 }  \over {
  c^2 l}}$ restoring the equation of conservation of energy.

Note that in the $x$ configuration of charges, the work of
the radiation field vanishes because in this case the radiation fields at the
locations of the particles vanish.

\section{RESOLUTION OF THE PARADOX I}
\label{PARI}

Now we have learned all the effects necessary for the resolution of  Paradox
I. In fact, we have seen a few effects which do  not play a role
in this case, but understanding them increases our confidence that our
explanation is the correct one.

 The anomalous transformation of the
electromagnetic energy is not relevant because the charges are at rest
at the beginning and at the end of the process. The subtle effect of
the work performed by the radiation field is not present too, since
the radiation fields in the locations of the particles vanish. It is
the interference of radiation which we  have not taken into account
that resolves the paradox. The
radiation energy (\ref{rad})  is much larger than the
term (\ref{gain}) which we have to compensate,  see (\ref{2term<}). However, we would 
like to get quantitative resolution of this paradox showing how  the
missing term (\ref{gain}) arises from the calculation of the radiation energy.

In order to obtain a quantitative result we specify how we perform the
process described in Section I. We assume that we perform it exactly
as in all other setups described here: we accelerate particles during
small time $\tau$ until they reach velocity $v$. In case (i), the
particle moves with constant velocity the distance $x$ when it stops
in the same manner as it was accelerated. In case (ii), both particles
reach velocity $v$ (absolute value) and stop at time $t$ after passing
the distance $x/2$. The time $t$ is short enough such that  each particle cannot receive a signal during
its motion about the motion of the
other particle. Thus,
\begin{equation}
\label{param}
\tau \ll t = {x \over {2v}} <
 {l  \over  {c+v}}. 
\end{equation}

In case (i) the radiation energy is created in the same amount due to
the acceleration and due to the stopping of the particle and, therefore, it is
twice the amount given by the Larmor formula (\ref{rad}):
\begin{equation}
\label{radi}
 R^{\rm i} = {4 \over 3}
 {{q^2 v^2}  \over { c^3\tau}}. 
\end{equation}

In case (ii) there are four events of changing velocity of a particle
by amount $v$ and, therefore, there are four spherical shells of
radiation field of the width $\tau c$, see Fig.~5. The radiation
energy is four times the Larmor energy (\ref{rad}) with the correction
due to the interference. The correction due to the interference has
four terms. The interferences are due to radiation emitted during the acceleration of the two
particles, stopping of the two particles, acceleration of the first
and stopping of the second, and acceleration of the second and stopping
of the first. All these terms can be calculated in the same way as we
have calculated the interference of radiation energy of two stopping
charged particles in Section IV.

  Accelerations and decelerations of the
particles are performed simultaneously, therefore, the direction of
interference is $\theta = {\pi\over 2}$.  This is the direction of the
maximal power of the radiation energy, see (\ref{radS}). The range of
the angles for which there is the interference is given by
(\ref{thetaD}) and, thus, similarly to derivation of (\ref{Eradint}),
we obtain that the interference term due to simultaneous acceleration
is equal to the term due to simultaneous stopping, and it equals
 \begin{equation}
  \label{aass}
   - {{q^2 v^2 }  \over { c^2 l}},
\end{equation}
 where the minus sign is because particles accelerate in opposite
 directions and the second term of (\ref{Eradint}) does not appear
 since simultaneity corresponds to $x=0$ in the notation of Section III.

 The interference between acceleration of the first and stopping of
the second takes place in the direction $\hat \theta_1$ defined by
\begin{equation}
\label{theta1}
\sin (\theta_1 -{\pi\over2}) ={{c t} \over{ l-vt}}  .
\end{equation}  
 For calculating this correction we can use (\ref{Eradint}) again,
taking into account that the particle stops after passing the distance
$vt={x\over2}$.  Therefore, this contribution is:
 \begin{equation}
\label{cont2}
  {{q^2 v^2 }  \over { c^2 l}} - 
 {{q^2 x^2}\over {4l^3}} .
\end{equation}

The contribution to the
correction of the radiation energy due to the interference
between acceleration of the second and stopping of the first particles
takes place in the direction defined by
\begin{equation}
\label{theta2}
\sin (\theta_2 -{\pi\over2}) ={{c t} \over{ l+vt}}.
\end{equation} 

\vskip .4cm
\begin{center} \leavevmode \epsfbox{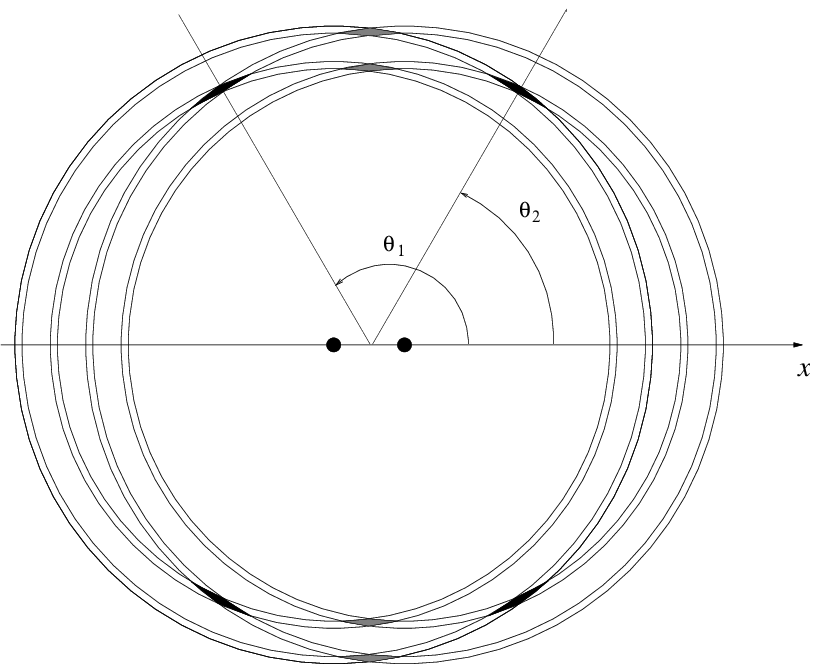} \end{center}

\noindent 
{\small {\bf Fig.~5.} Electromagnetic radiation of the two charged
  particles which are simultaneously accelerated toward each other
  and after time $t$ stopped, case (ii). The shadowed area signifies
  destructive interference and the area painted in black signifies
  constructive interference.}

\noindent
In our case $l\gg vt$, so we can make an approximation 
$l-vt~\approx l +vt \approx l$ and, therefore, we get the same expression again. Summing
up all the expressions, we obtain:
\begin{equation}
\label{radii}
 R^{\rm ii} = {8 \over 3} {{q^2 v^2}  \over { c^3\tau}} -
 {{2q^2 v^2 }  \over { c^2 l}}+
2\left( {{q^2 v^2 }  \over { c^2 l}} - 
 {{q^2 x^2}\over {4l^3}}\right)= {8 \over 3}
 {{q^2 v^2}  \over { c^3\tau}}  - 
 {{q^2 x^2}\over {2l^3}} . 
\end{equation}

 Now we are able to analyze the setup of Paradox I taking into account 
 the radiation energy. In case (i) the work performed by the external 
 forces should include the radiation energy (\ref{radi}). Thus,
 instead of  (\ref{W}) we obtain 
\begin{equation}
\label{W+rad}
W^{\rm i} = U_{NEW} - U_{OLD}+  R^{\rm i} = {q^2 \over {l-x}} - 
  {q^2 \over {l}} + {4 \over 3}
 {{q^2 v^2}  \over { c^3\tau}}.
\end{equation}

In case (ii), following the structure of Paradox I, we have to
calculate the work taking into account the causality argument: each
particle ``does not know'' that the other particle moved. Therefore,
the work against the field and the radiated energy should be
calculated as if the other particle has not moved. The work is
twice the amount of work in case (i) with the change of $x \rightarrow
{x\over2}$. Thus, instead of (\ref{W'}), we obtain
\begin{equation}
W^{\rm ii} = W_1 + W_2 =2 \left( {q^2 \over {l-{x\over 2}}} -   {q^2 \over l} + {4 \over 3}
 {{q^2 v^2}  \over { c^3\tau}}\right) .
\label{W'+rad}
\end{equation}
    
Clearly we cannot gain energy from constructing a machine with a
    cycle of process (ii) and reversed process (i).
The work required for reversed process (i) is: 
\begin{equation}
\label{w+radneg}
W^{\rm \tilde i}  = 
  {q^2 \over {l}}- {q^2 \over {l-x}} + {4 \over 3}
 {{q^2 v^2}  \over { c^3\tau}}.
\end{equation}
Thus, the work during the whole cycle is: 
\begin{eqnarray}
\nonumber
W_{tot} = W^{\rm \tilde i}+W^{\rm ii}={q^2 \over l}- {q^2 \over {l-x}} + {4 \over 3}
 {{q^2 v^2}  \over { c^3\tau}}+~~~~~~~\\ ~~~~~~~2\left( {q^2 \over {l-{x\over 2}}} -   {q^2 \over l} 
+ {4 \over 3}
 {{q^2 v^2}  \over { c^3\tau}}\right) \approx - {{q^2 x^2} \over {2l^3}} +
 {{4q^2 v^2}  \over { c^3\tau}}.
\label{Wtot}
\end{eqnarray}
This work is greater than zero, since the radiation term is much larger than the gain in the potential energy; it can be seen explicitly using (\ref{param}).
 However, even if we collect the radiation energy, we still
    cannot gain energy. Indeed, the total radiation energy is: 
\begin{equation}
\label{radtot}
 R_{tot}= R^{\rm i}+ R^{\rm ii} 
=  {4 \over 3}
 {{q^2 v^2}  \over { c^3\tau}}+{8 \over 3}
 {{q^2 v^2}  \over { c^3\tau}}  - 
 {{q^2 x^2}\over {2l^3}}=  
 {{4q^2 v^2}  \over { c^3\tau}}  - 
 {{q^2 x^2}\over {2l^3}} . 
\end{equation}
We obtained exactly the same expression, i.e. our calculations have  shown 
(up to the precision of the order of ${v^2\over c^2}$) that during the
complete cycle $ W_{tot}= R_{tot}$.
This completes the analysis of the paradox presented at the beginning
of this paper. 

Is it a simple task to demonstrate conservation of energy to a higher
order in $v\over c$? It is not difficult to expand the algebraic
expressions we have to a higher order, but this is not enough. We have
used more approximations, in particular, the formulas for the
radiation of the charged particles we have used are correct only in
the approximation of small acceleration and small velocities. Indeed,
Eq. (\ref{rad}) cannot be universally correct, since it says that by
reducing $\tau$, the time of stopping the charged particle, we can
obtain unlimited amount of radiation energy: clearly we cannot get
more energy than the particle has. Thus, higher order calculations of
the equation of conservation of energy is an elaborate task which goes
beyond the scope of this paper.

In this paper we have analyzed some relativistic features of the classical
electromagnetic theory. We demonstrated in a quantitative way the
relevance of the several effects to the balance of conservation of
energy in a system of charges. These effects are: changing of the
electromagnetic field according to relativistic causality constraints,
anomalous transformation of the electromagnetic energy, energy
radiated by accelerated charges, interference of radiation, and work
performed by the radiation field. 
 We believe that presenting the
subject in the form of ``paradoxes'' helps to achieve a deeper
understanding. Obtaining quantitative resolutions of 
presented paradoxical situations helps to reach  confidence
in  
applying  the equation of conservation of energy  for indirect
calculations of various effects.

\vspace{.3cm}
 \centerline{\bf  ACKNOWLEDGMENTS}
 
 It is a pleasure to thank Shmuel Nussinov, 
 and Philip Pearle for helpful discussions.
 This research was supported in part by 
the EPSRC grant  GR/N33058.


\end{document}